\documentclass[aps,preprint,amsmath,tightenlines,showpacs,showkeys]{revtex4}

\usepackage{bm}

\begin{document}

\title{Statistical comparison of quark mass matrices\\ in the physical basis}

\author{S.~Chaturvedi}

\email[]{scsp@uohyd.ernet.in}

\affiliation{
School of Physics.
University of Hyderabad.\\
Hyderabad 500 046. INDIA.
}

\author{V.~Gupta}

\email[]{virendra@mda.cinvestav.mx}

\affiliation{
Departamento de F\'{\i}sica Aplicada.\\
Centro de Investigaci\'on y de Estudios Avanzados del IPN.\\
Unidad M\'erida.\\
A.P. 73, Cordemex.\\
M\'erida, Yucat\'an, 97310. MEXICO.
}

\author{G.~S\'anchez-Col\'on}

\email[]{gsanchez@mda.cinvestav.mx}

\affiliation{
Departamento de F\'{\i}sica Aplicada.\\
Centro de Investigaci\'on y de Estudios Avanzados del IPN.\\
Unidad M\'erida.\\
A.P. 73, Cordemex.\\
M\'erida, Yucat\'an, 97310. MEXICO.
}

\author{S.~Rajpoot}

\email[]{rajpoot@csulb.edu}

\affiliation{
Department of Physics \& Astronomy.\\
California State University, Long Beach.\\
Long Beach, CA 90840. USA.
}

\date{\today}

\begin{abstract}

Using the four best measured moduli of the flavor mixing matrix ($|V_{\rm ud}|$,
$|V_{\rm us}|$, $|V_{\rm cd}|$, $|V_{\rm cs}|$), the Jarlskog invariant $J(V)$,
and the quark masses at $M_Z$ energy scale as experimental constraints, a
statistical comparison of three different types of quark mass matrices in the
physical basis is performed. The mass matrices in question are the
Chaturvedi-Gupta-S\'anchez-Col\'on (CGS), the Fritzsch  and the Gupta-Rajpoot
types. With nine parameters the best fits are obtained using a Gupta-Rajpoot
type matrix while with seven parameters the best fits are obtained using the
CGS type matrix. The stability of our analysis with respect to evolution of the
quark masses is also presented.

\end{abstract}


\maketitle

\section{\label{introduction}Introduction}
An important aspect of the standard model is the understanding of the structure of quark mass matrices that adequately describe the observed quark flavor mixing. The hope is that such studies will unravel the underlying symmetries responsible for the required texture in such mass matrices. Deviations form the symmetries will provide important constraints on possible new physics beyond the standard model.

In general, the $3\times 3$ hermitian quark mass matrices $M_{\rm u}$ and
$M_{\rm d}$ have nine real parameters each. These determine the six quark
masses and the four independent real parameters in the quark mixing matrix $V$.
Unfortunately, up to now one has no guiding principle for choosing $M_{\rm u}$
and $M_{\rm d}$ with fewer parameters. Recently~\cite{chaturvedi08,gupta09},
constraints on three type of quark mass matrices were obtained using the
experimentally determined values of the quark masses and measured properties of
the mixing matrix $V$. In this paper we extend the analysis of
Refs.~\cite{chaturvedi08,gupta09} to a different choice of mass matrices in the
physical basis, namely, Fritzsch~\cite{fritzsch70,fritzsch78,fritzsch86},
Gupta-Rajpoot~\cite{gupta91,rajpoot92}, and CGS~\cite{chaturvedi08,gupta09}.

Fritzsch~\cite{fritzsch70,fritzsch78,fritzsch86} suggested use of $M_{\rm u}$
and $M_{\rm d}$ with three textures ({\it viz.} $M_{11}=M_{22}=M_{13}=0$). In
this case, each quark mass matrix has five real parameters in general. The
Gupta-Rajpoot~\cite{gupta91,rajpoot92} mass matrix is similar to the Fritzsch
type with one less texture, namely $M_{22}\ne 0$. Each Gupta-Rajpoot type of
mass matrix has six real parameters in general. Lastly, we consider the
CGS~\cite{chaturvedi08,gupta09} type matrix suggested recently. This has only
two textures, $M_{11}=M_{22}=0$. For $M_{13}=0$ this reduces to the Fritzsch
type matrix. CGS type matrix has the virtue that it will give CP-violation
in all the three bases (physical, $M_{\rm u}$ diagonal or $M_{\rm d}$ diagonal)
because it can have ${\rm Im}(M_{12}\,M_{23}\,M^{*}_{13}) \ne
0$~\cite{chaturvedi08}. Note that Fritzsch and Gupta-Rajpoot type matrices will
give CP-violation only in the physical basis. In general, each CGS type of
matrix has seven parameters. However, a unitary transformation relates both
$M_{\rm u}$ and $M_{\rm d}$ of CGS type to, for example, $M_{\rm u}$ Fritzsch
type and $M_{\rm d}$ CGS type.

Section~\ref{notation} gives notation and basic formulas or expressions needed
and the general procedure adopted for the analysis. Confrontation with
experimental data for each type of matrices is presented in
Secs.~\ref{sectionfritzsch}, \ref{sectiongr}, and \ref{sectioncgs}, were we
consider in the physical basis the Fritzsch, the Gupta-Rajpoot, and the CGS
types of mass matrices, respectively. Results obtained for the various
cases are compared and discussed in the final Sec.~\ref{conclusions}.

\section{\label{notation}Notation and basic formulas}

The $3\times 3$ hermitian quark mass matrix $M_{\rm q}$ is diagonalized
by $V_{\rm q}$ so that $M_{\rm q}=V_{\rm q}^{\dagger}\hat{M_{\rm
q}}V_{\rm q}\,$, q=u,d. The eigenvalues are denoted by ($\lambda_{\rm
u}$,$\lambda_{\rm c}$,$\lambda_{\rm t}$) and ($\lambda_{\rm
d}$,$\lambda_{\rm s}$,$\lambda_{\rm b}$) for the up and down quark mass
matrices, respectively. Note that the eigenvalues are real but not
necessarily positive. Each mass matrix can be expressed in terms of its
projectors. Thus,

\begin{equation}
M_{\rm u}=\sum_{\alpha={\rm u, c, t}}\lambda_{\alpha}N_{\alpha}
\qquad
{\rm and}
\qquad
M_{\rm d}=\sum_{j={\rm d, s, b}}\lambda_{j}N_{j}.
\label{projectors}
\end{equation}

\noindent
Since $V= V_{\rm u}V^\dagger_{\rm d}$, it follows that~\cite{18}

\begin{equation}
|V_{\alpha\, j}|^2={\rm Tr}[N_\alpha N_j],
\label{5}
\end{equation}

\noindent
where

\begin{equation}
N_\alpha= \frac{(\lambda_\beta-M_{\rm u})(\lambda_\gamma
-M_{\rm u})}{(\lambda_\beta-\lambda_\alpha) (\lambda_\gamma
-\lambda_\alpha)}
\label{4}
\end{equation}

\noindent
and

\begin{equation}
N_j=
\frac{(\lambda_k-M_{\rm d})(\lambda_l -M_{\rm
d})}{(\lambda_k-\lambda_j) (\lambda_l -\lambda_j)},
\label{4p}
\end{equation}

\noindent
with ($\alpha$,$\beta$,$\gamma$) and ($j$,$k$,$l$) any permutation of
(u,c,t) and (d,s,b), respectively. By unitarity only four of the nine
$|V_{\alpha\,j}|^2$ are independent.

The Jarlskog invariant $J(V)$, which is a measure of CP-violation can be
directly expressed in terms of $M_{\rm u}$ and $M_{\rm d}$ and their
eigenvalues~\cite{4}, thus

\begin{equation}
{\rm Det}([M_{\rm u},M_{\rm d}])=2i\,D(\lambda_\alpha)D(\lambda_j) J(V),
\label{12}
\end{equation}

\noindent
where $D(\lambda_\alpha)=
(\lambda_{\rm c}-\lambda_{\rm u})(\lambda_{\rm t}-\lambda_{\rm u})
(\lambda_{\rm t}-\lambda_{\rm c})$
and
$
D(\lambda_j)=
(\lambda_{\rm s}-\lambda_{\rm d})(\lambda_{\rm b}-\lambda_{\rm d})
(\lambda_{\rm b}-\lambda_{\rm s})$.

For a given choice of the mass matrices texture, most of the elements of the
quark mass matrices can be expressed in terms of the mass eigenvalues (the free
parameters) using the characteristic equations for $M_{\rm u}$ and $M_{\rm d}$
and others have to be considered as supplemental free parameters. We perform a
$\chi^2$ analysis with a single $\chi^2$ function formed by eleven
constraints~\cite{pdg06} $|V^{\rm exp}_{\rm ud}|$, $|V^{\rm exp}_{\rm us}|$,
$|V^{\rm exp}_{\rm cd}|$, $|V^{\rm exp}_{\rm cs}|$, $J^{\rm exp}(V)$, and
$\lambda^{\rm exp}_{\rm q}$ $(=\pm m^{\rm exp}_{\rm q},\ {\rm q=u,c,t,d,s,b})$,
the eigenvalues of the quark mass matrices at a specified energy
scale~\cite{xing08}. This function is

\begin{eqnarray} \chi^2 &=& \left(\frac{|V^{\rm exp}_{\rm ud}| - |V_{\rm
ud}|}{\sigma_{|V_{\rm ud}|}}\right)^2 + \left(\frac{|V^{\rm exp}_{\rm us}| -|
V_{\rm us}|}{\sigma_{|V_{\rm us}|}}\right)^2 + \left(\frac{|V^{\rm exp}_{\rm
cd}| - |V_{\rm cd}|}{\sigma_{|V_{\rm cd}|}}\right)^2 + \left(\frac{|V^{\rm
exp}_{\rm cs}| - |V_{\rm cs}|}{\sigma_{|V_{\rm cs}|}}\right)^2 + \nonumber \\ &
& + \left(\frac{J^{\rm exp}(V)-J(V)}{\sigma_{J(V)}}\right)^2 + \sum_{{\rm q = u,
c, t, d, s, b}}\left(\frac{\lambda^{\rm exp}_{\rm q}-\lambda_{\rm q}}
{\sigma_{\lambda_{\rm q}}}\right)^2. \label{chi2} \end{eqnarray}

\noindent
This is an over constrained system of restrictions for the free
parameters $\lambda_{\rm q}$ (q=u,c,t,d,s,b) and the supplemental free
parameters for each texture if it is the case.

The numerical values~\cite{pdg06,xing08} of the eleven experimental constraints
used (the six quark masses, the four moduli from the mixing matrix ($|V_{\rm
ud}|$, etc.), and $J(V)$) are given in Table~\ref{table1} for easy reference.

\section{\label{sectionfritzsch}Fritzsch type mass matrices}

Fritzsch type mass
matrices~\cite{fritzsch70,fritzsch78,fritzsch86} are given by
the hermitian matrices

\begin{equation}
M_{\rm u}=
\begin{pmatrix}
0 & A & 0 \\
A^* & 0 & B \\
0 & B^* & C
\end{pmatrix}
,
\quad\quad
M_{\rm d}=
\begin{pmatrix}
0 & A' & 0 \\
A'^* & 0 & B' \\
0 & B'^* & C'
\end{pmatrix}.
\label{fritzsch}
\end{equation}

\noindent
Without lack of generality we can take $C$ and $C'$ to be positive and
$A$ and $B$ to be real and positive. Then, $M_{\rm u}$ and $M_{\rm d}$ have
eight real parameters $A$, $B$, $C$, $C'$, $|A'|$, $|B'|$ and the phases
$\phi_{A'}$ and $\phi_{B'}$. This type of of mass matrices can only be used in
the physical basis because an off-diagonal matrix element is zero.

From the characteristic polynomial of $M_{\rm u}$
and $M_{\rm d}$ the six parameters $A$, $B$, $C$, $C'$,
$|A'|$, and $|B'|$ are expressed in terms of the eigenvalues

\[
A=\left[
-\frac{\lambda_{\rm u}\lambda_{\rm c}\lambda_{\rm t}}
{\lambda_{\rm u}+\lambda_{\rm c}+\lambda_{\rm t}
}
\right]^{1/2}
,
\qquad
C=\lambda_{\rm u}+\lambda_{\rm c}+\lambda_{\rm t},
\]

\begin{equation}
B=\left[-
\frac{
(\lambda_{\rm t}+\lambda_{\rm c})
(\lambda_{\rm t}+\lambda_{\rm u})
(\lambda_{\rm c}+\lambda_{\rm u})}
{\lambda_{\rm u}+\lambda_{\rm c}+\lambda_{\rm t}
}
\right]^{1/2}
,
\label{15b}
\end{equation}

\noindent
and similarly for $M_{\rm d}$. The parameters $|A'|$, $|B'|$, and $C'$
are obtained by replacing ($A$,$B$,$C$) by ($|A'|$,$|B'|$,$C'$) and
($\lambda_{\rm u}$,$\lambda_{\rm c}$,$\lambda_{\rm t}$) by ($\lambda_{\rm
d}$,$\lambda_{\rm s}$,$\lambda_{\rm b}$).

According to~(\ref{5}), the theoretical expressions for the magnitudes of the
unitary quark mixing matrix elements in this case are given by

\begin{eqnarray}
|V_{\alpha\,j}|^2&=&
\big{[}
(\lambda_{\alpha} - \lambda_{\beta})
(\lambda_{\alpha} - \lambda_{\gamma})
(\lambda_j - \lambda_k)
(\lambda_j - \lambda_l)
\big{]}^{-1}\times
\nonumber
\\
&\ &
\bigg\{
\big{(}\lambda_{\beta}\lambda_{\gamma} + A^2 + B^2\big{)}
\big{(}\lambda_k\lambda_l + {|A'|}^2 + {|B'|}^2\big{)}
+
\big{(}\lambda_{\beta}\lambda_{\gamma} + A^2\big{)}
\big{(}\lambda_k\lambda_l + {|A'|}^2\big{)}
\nonumber
\\
&\ &+\,
\Big{[}
\left(\lambda_{\alpha} + \lambda_{\beta}\right)
\left(\lambda_{\alpha} + \lambda_{\gamma}\right)
+ B^2
\Big{]}
\left[
\left(\lambda_j + \lambda_k\right)
\left(\lambda_j + \lambda_l\right)
+ {|B'|}^2
\right]
\nonumber
\\
&\ &+\,
2
\left(\lambda_{\beta} + \lambda_{\gamma}\right)
\left(\lambda_k + \lambda_l)\,
A{|A'|}\cos(\phi_{A'}\right)
\nonumber
\\
&\ &+\,
2\lambda_{\alpha}\lambda_j\,B{|B'|}\cos(\phi_{B'}) +
2B\,A{|B'|}{|A'|}\cos(\phi_{A'} + \phi_{B'})
\bigg\}\,,
\label{vaj}
\end{eqnarray}

\noindent
where ($\alpha$,$\beta$,$\gamma$) is any permutation of (u,c,t) and
($j$,$k$,$l$) any permutation of (d,s,b).

In this case, the theoretical expression for the Jarslkog invariant $J(V)$ given
by Eq.~(\ref{12}) translates into

\begin{eqnarray}
J(V)&=&
\big{[}
(\lambda_{\rm t} - \lambda_{\rm c}) (\lambda_{\rm t} -
\lambda_{\rm u}) (\lambda_{\rm c} - \lambda_{\rm u})(\lambda_{\rm b} -
\lambda_{\rm s}) (\lambda_{\rm b} - \lambda_{\rm d})(\lambda_{\rm s} -
\lambda_{\rm d})
\big{]}^{-1} \times
\nonumber
\\ &\ &
\bigg\{
\Big[
A\,{|A'|}\sin(\phi_{A'}) - B\,{|B'|}\sin(\phi_{B'})
\Big]
\nonumber
\\ &\ & \quad \times
\Big[
A^2{|B'|}^2+B^2 {|A'|}^2-2\,A\,B\,{|A'|}\,{|B'|}\cos(\phi_{A'}+\phi_{B'})
\Big]
\nonumber
\\ &\ & \quad - \,
A\,{|A'|}\sin(\phi_{A'})
\Big[
C^2 {|B'|}^2 + B^2{C'}^2 - 2\,C\,B\,{C'}\,{|B'|}\cos(\phi_{B'})
\Big]
\bigg\}\,.
\label{jv}
\end{eqnarray}

The mass matrices $M_{\rm u}$ and $M_{\rm d}$ in Eq.~(\ref{fritzsch}) do not
have positive definite eigenvalues. Thus, $\lambda^2_{\rm u}=m^2_{\rm u}$,
$\lambda^2_{\rm d}=m^2_{\rm d}$, etc., where $m_{\rm u}$ is the (positive) mass
of the up quark, etc. Using the mass hierarchies
$|\lambda_{\rm u}|<<|\lambda_{\rm c}|<<|\lambda_{\rm t}|$ for the up
quark sector and $|\lambda_{\rm d}|<<|\lambda_{\rm s}|<<|\lambda_{\rm
b}|$ for the down one and the characteristic equations it is possible to fix
the relative phases between the eigenvalues and the quark
masses~\cite{chaturvedi08},

\begin{equation}
(\lambda_{\rm u}\,,\lambda_{\rm c}\,,\lambda_{\rm t}) =
(m_{\rm u}\,,-m_{\rm c}\,,m_{\rm t})
\quad
{\rm and}
\quad
(\lambda_{\rm d}\,,\lambda_{\rm s}\,,\lambda_{\rm b}) =
(m_{\rm d}\,,-m_{\rm s}\,,m_{\rm b})\,.
\label{17p}
\end{equation}

For this texture there are eight free parameters to be estimated, the six
eigenvalues ($\lambda_{\rm u}$,$\lambda_{\rm c}$,$\lambda_{\rm t}$) and
($\lambda_{\rm d}$,$\lambda_{\rm s}$,$\lambda_{\rm b}$), and the two phases
$\phi_{A'}$ and $\phi_{B'}$. The number of degrees of freedom is three (eleven
constraints with eight free parameters). At the $M_Z$ energy scale the fitted
values for the eigenvalues are $\lambda_{\rm u} = (1.66^{+0.40}_{-0.36})\,{\rm
MeV}$, $\lambda_{\rm c} = -(0.621\pm 0.080)\,{\rm GeV}$, $\lambda_{\rm t} =
(172.4\pm 3.0)\,{\rm GeV}$, $\lambda_{\rm d} = (2.29^{+0.50}_{-0.49})\,{\rm
MeV}$, $\lambda_{\rm s} = -(38.0^{+4.3}_{-5.8})\,{\rm MeV}$, $\lambda_{\rm b} =
(2.901\pm 0.090)\,{\rm GeV}$, and the fitted values for the phases are
$\phi_{A'} = (-71^{+19}_{-23})^\circ$ and $\phi_{B'} = (-4^{+18}_{-16})^\circ$.
The total $\chi^2/({\rm dof}) = 4.23/3=1.41$.

We observe that the fitted values obtained for the phases $\phi_{A'}$ and
$\phi_{B'}$ are compatible with $-\pi/2$ and $0$, respectively. The
corresponding results obtained for this particular choice in this six free
parameters fit (the six eigenvalues) and five degrees of freedom at the same
$M_Z$ energy scale are: $\lambda_{\rm u} = (1.66^{+0.40}_{-0.36})\,{\rm MeV}$,
$\lambda_{\rm c} = -(0.622\pm 0.079)\,{\rm GeV}$, $\lambda_{\rm t} = (172.4\pm
3.0)\,{\rm GeV}$, $\lambda_{\rm d} = (1.98\pm 0.22)\,{\rm MeV}$, $\lambda_{\rm
s} = -(38.3^{+4.0}_{-3.9})\,{\rm MeV}$, $\lambda_{\rm b} = (2.902\pm
0.090)\,{\rm GeV}$. The total $\chi^2/({\rm dof}) = 4.84/5=0.97$.

It is interesting to note that the last fit was obtained with only six real
parameters because $M_{\rm u}$ is real and in $M_{\rm d}$, $B'$ and $C'$ are
real and $A'$ is pure imaginary.

\section{\label{sectiongr}Gupta-Rajpoot type mass matrices}

For the three families of quarks the Gupta-Rajpoot mass matrices
are of the form~\cite{gupta91,rajpoot92}

\begin{equation}
M_{\rm u}=
\begin{pmatrix}
0 & A & 0 \\
A^* & |D| & B \\
0 & B^* & |C|
\end{pmatrix}
,
\quad\quad
M_{\rm d}=
\begin{pmatrix}
0 & A' & 0 \\
A'^* & |D'| & B' \\
0 & B'^* & |C'|
\end{pmatrix}
.
\label{guptarajpoot}
\end{equation}

\noindent
In their general form, these matrices have been extensively studied in the
literature for both quarks mixing and lepton mixing~\cite{rajpoot07}. Without
lack of generality we can make $M_{\rm u}$ to be real and positive. Then,
$M_{\rm u}$ and $M_{\rm d}$ have ten real parameters $A$, $B$, $C$, $D$,
$|A'|$, $|B'|$, $|C'|$, $|D'|$, and the phases $\phi_{A'}$ and $\phi_{B'}$.

From the characteristic equations the parameters $A$, $B$,
$C$, $|A'|$, $|B'|$, and $|C'|$ are expressed in terms of the six
eigenvalues and $D$. For $M_{\rm u}$,

\[
A=\left[
-\frac{\lambda_{\rm u}\lambda_{\rm c}\lambda_{\rm t}}
{\lambda_{\rm u}+\lambda_{\rm c}+\lambda_{\rm t}-D
}
\right]^{1/2}
,
\qquad
C=\lambda_{\rm u}+\lambda_{\rm c}+\lambda_{\rm t}-D,
\]

\begin{eqnarray}
B&=&
\left(
\lambda_{\rm u}+\lambda_{\rm c}+\lambda_{\rm t}-D
\right)^{-1/2}
\times
\nonumber \\
&\ &
\bigg\{
-
(\lambda_{\rm t}+\lambda_{\rm c})
(\lambda_{\rm t}+\lambda_{\rm u})
(\lambda_{\rm c}+\lambda_{\rm u})
\nonumber \\
&\ &+\,
\Big{[}
(\lambda_{\rm t}+\lambda_{\rm c})
(\lambda_{\rm t}+\lambda_{\rm u})
+
(\lambda_{\rm t}+\lambda_{\rm u})
(\lambda_{\rm c}+\lambda_{\rm u})
+
(\lambda_{\rm c}+\lambda_{\rm u})
(\lambda_{\rm t}+\lambda_{\rm c})
\Big{]} D
\nonumber \\
&\ &-\,
2\,
(\lambda_{\rm u}+\lambda_{\rm c}+\lambda_{\rm t}) D^2
+
D^3
\bigg\}^{1/2}\,,
\label{15bgr}
\end{eqnarray}

\noindent
and similarly for $M_{\rm d}$ where the parameters $|A'|$, $|B'|$, and $|C'|$
are obtained by replacing ($A$,$B$,$C$) by ($|A'|$,$|B'|$,$|C'|$),
($\lambda_{\rm u}$,$\lambda_{\rm c}$,$\lambda_{\rm t}$) by ($\lambda_{\rm
d}$,$\lambda_{\rm s}$,$\lambda_{\rm b}$), and $D$ by $|D'|$.

According to Eq.~(\ref{5}), the theoretical expressions for the observable
magnitudes of the unitary quark mixing matrix elements are given in this case
by

\begin{eqnarray}
|V_{\alpha\,j}|^2&=&
\big{[}
(\lambda_{\alpha} - \lambda_{\beta})
(\lambda_{\alpha} - \lambda_{\gamma})
(\lambda_j - \lambda_k)
(\lambda_j - \lambda_l)
\big{]}^{-1}
\times
\nonumber \\
&\ &
\bigg\{
\big{(}
A^2 + \lambda_{\beta}\lambda_{\gamma}
\big{)}
\big{(}
{|A'|}^2 + \lambda_k\lambda_l
\big{)}
\nonumber \\
&\ &+\,
\Big{[}
B^2 +
\left(\lambda_{\alpha} + \lambda_{\beta} - D\right)
\left(\lambda_{\alpha} + \lambda_{\gamma} - D\right)
\Big{]}
\Big{[}
{|B'|}^2 +
\left(\lambda_j + \lambda_k - |D'|\right)
\left(\lambda_j + \lambda_l - |D'|\right)
\Big{]}
\nonumber \\
&\ &+\,
\Big{[}A^2 + B^2 +
\big{(}\lambda_{\beta}-D\big{)}\big{(}\lambda_{\gamma}-D\big{)} \Big{]}
\Big{[}{|A'|}^2 + {|B'|}^2 +
\big{(}\lambda_k-|D'|\big{)}\big{(}\lambda_l-|D'|\big{)} \Big{]}
\nonumber \\
&\ &+\,
2
\left(\lambda_{\beta} + \lambda_{\gamma}-D\right)
\left(\lambda_k + \lambda_l-|D'|)\,
A{|A'|}\cos(\phi_{A'}\right)
\nonumber \\
&\ &+\,
2
\lambda_{\alpha}\lambda_j\,B{|B'|}\cos(\phi_{B'}) +
2
A\,B{|A'|}{|B'|}\cos(\phi_{A'} + \phi_{B'})
\bigg\}\,,
\label{vajgr}
\end{eqnarray}

\noindent
where ($\alpha$,$\beta$,$\gamma$) is any permutation of (u,c,t) and
($j$,$k$,$l$) any permutation of (d,s,b).

The theoretical expression for the observable Jarslkog invariant $J(V)$ for
this case is obtained from Eq.~(\ref{12}),

\begin{eqnarray}
J(V)&=&
\big{[}
(\lambda_{\rm t} - \lambda_{\rm c}) (\lambda_{\rm t} -
\lambda_{\rm u}) (\lambda_{\rm c} - \lambda_{\rm u})(\lambda_{\rm b} -
\lambda_{\rm s}) (\lambda_{\rm b} - \lambda_{\rm d})(\lambda_{\rm s} -
\lambda_{\rm d})
\big{]}^{-1} \times
\nonumber
\\ & &
\bigg\{
\Big[
A\,{|A'|}\sin(\phi_{A'}) - B\,{|B'|}\sin(\phi_{B'})
\Big]
\nonumber
\\ &\ & \quad \times
\Big[
A^2{|B'|}^2+B^2 {|A'|}^2-2\,A\,B\,{|A'|}\,{|B'|}\cos(\phi_{A'}+\phi_{B'})
\Big]
\\ & & \, - \,
A\,{|A'|}\sin(\phi_{A'})
\Big[
C^2 {|B'|}^2 + B^2{C'}^2 - 2\,C\,B\,{C'}\,{|B'|}\cos(\phi_{B'})
\Big]
\nonumber
\\ & & \, + \,
A\,B\,{|D'|}
\Big[
B\,{|A'|}\,{C'}\sin(\phi_{A'}) + A\,{|B'|}\,{C'}\sin(\phi_{B'})
- C\,{|A'|}\,{|B'|}\,\sin(\phi_{A'} + \phi_{B'})
\Big]
\nonumber
\\ & & \, + \,
D\,{|A'|}\,{|B'|}
\Big[
A\,C\,{|B'|}\sin(\phi_{A'}) + B\,C\,{|A'|}\sin(\phi_{B'})
- A\,B\,{C'}\,\sin(\phi_{A'} + \phi_{B'})
\Big]
\bigg\}\,.
\nonumber
\label{jvgr}
\end{eqnarray}

To reduce the number of free parameters to nine we shall take $|D'|=0$. As
mentioned before, $\lambda^2_{\rm u}=m^2_{\rm u}$, $\lambda^2_{\rm d}=m^2_{\rm
d}$, etc., and we need a solution with the mass hierarchies $|\lambda_{\rm
u}|<<|\lambda_{\rm c}|<<|\lambda_{\rm t}|$ for the up quark sector and
$|\lambda_{\rm d}|<<|\lambda_{\rm s}|<<|\lambda_{\rm b}|$ for the down one. In
this case the relative signs between the $\lambda$'s and the quark masses can
not be fixed {\it a priori} and all the possible combinations coming from the
two signs in front of each of the quark masses were explored in the $\chi^2$
analysis. The result of this exploration is that the best fit for the
Gupta-Rajpoot case is obtained for the identification, Eq.~(\ref{17p}), of the
relative signs between $\lambda$'s and $m$'s.

For this texture, the nine free parameters estimated were the six eigenvalues
($\lambda_{\rm u}$,$\lambda_{\rm c}$,$\lambda_{\rm t}$) and ($\lambda_{\rm
d}$,$\lambda_{\rm s}$,$\lambda_{\rm b}$), the matrix element $D$, and the two
phases $\phi_{A'}$ and $\phi_{B'}$. The number of degrees of freedom is then
two. At the $M_Z$ energy scale the fitted values for the eigenvalues are
$\lambda_{\rm u} = (1.49^{+0.43}_{-0.40})\,{\rm MeV}$, $\lambda_{\rm c} =
-(0.610\pm 0.083)\,{\rm GeV}$, $\lambda_{\rm t} = (172.5\pm 3.0)\,{\rm GeV}$,
$\lambda_{\rm d} = (3.19^{+0.89}_{-0.80})\,{\rm MeV}$, $\lambda_{\rm s} =
-(57^{+14}_{-13})\,{\rm MeV}$, $\lambda_{\rm b} =
(2.888^{+0.089}_{-0.091})\,{\rm GeV}$, and the fitted values for the matrix
element $D$ and the phases are $D = (2.6^{+2.4}_{-2.0})\,{\rm GeV}$, $\phi_{A'}
= (-81^{+28}_{-27})^\circ$ and $\phi_{B'} = (-20^{+12}_{-4})^\circ$,
respectively. The total $\chi^2/({\rm dof}) = 0.80/2=0.40$.

The fitted values obtained for the phases $\phi_{A'}$ and $\phi_{B'}$ are still
compatible with $-\pi/2$ and $0$, respectively. The corresponding results
obtained for this particular choice in this seven free parameters fit (the six
eigenvalues and $D$) and four degrees of freedom at the same $M_Z$ energy scale
are: $\lambda_{\rm u} = (1.69^{+0.40}_{-0.36})\,{\rm MeV}$, $\lambda_{\rm c} =
-(0.591^{+0.082}_{-0.081})\,{\rm GeV}$, $\lambda_{\rm t} = (172.5\pm 3.0)\,{\rm
GeV}$, $\lambda_{\rm d} = (2.92\pm 0.69)\,{\rm MeV}$, $\lambda_{\rm s} = -(57\pm
13)\,{\rm MeV}$, $\lambda_{\rm b} = (2.889\pm 0.090)\,{\rm GeV}$, and $D =
(0.68^{+0.52}_{-0.48})\,{\rm GeV}$. The total $\chi^2/({\rm dof}) =
2.62/4=0.66$.

We note that the last fit was obtained with only seven free parameters since
$M_{\rm u}$ has four real parameters while $M_{\rm d}$ is Fritzsch type with
$A'$ pure imaginary.

Another option to reduce the number of free parameters to nine is with $D=0$ and
$|D'|\ne 0$. All the possible combinations coming from the two signs in front of
each of the quark masses were explored in the $\chi^2$ analysis and the best fit
is obtained for the identification,

\begin{equation}
(\lambda_{\rm u}\,,\lambda_{\rm c}\,,\lambda_{\rm t}) =
(m_{\rm u}\,,-m_{\rm c}\,,m_{\rm t})
\quad
{\rm and}
\quad
(\lambda_{\rm d}\,,\lambda_{\rm s}\,,\lambda_{\rm b}) =
(-m_{\rm d}\,,m_{\rm s}\,,-m_{\rm b})\,.
\label{17pp}
\end{equation}

\noindent
Now, the nine free parameters estimated were the six eigenvalues ($\lambda_{\rm
u}$,$\lambda_{\rm c}$,$\lambda_{\rm t}$) and ($\lambda_{\rm d}$,$\lambda_{\rm
s}$,$\lambda_{\rm b}$), the matrix element $|D'|$, and the two phases
$\phi_{A'}$ and $\phi_{B'}$. The number of degrees of freedom is two. At the
$M_Z$ energy scale the fitted values for the eigenvalues are $\lambda_{\rm u} =
(1.60^{+0.41}_{-0.37})\,{\rm MeV}$, $\lambda_{\rm c} = -(0.603\pm 0.083)\,{\rm
GeV}$, $\lambda_{\rm t} = (172.5\pm 3.0)\,{\rm GeV}$, $\lambda_{\rm d} =
-(3.13^{+0.86}_{-0.79})\,{\rm MeV}$, $\lambda_{\rm s} = (55^{+14}_{-13})\,{\rm
MeV}$, $\lambda_{\rm b} = -(2.889\pm 0.090)\,{\rm GeV}$, and the fitted values
for the matrix element $|D'|$ and the phases are $|D'| = (32^{+25}_{-23})\,{\rm
MeV}$, $\phi_{A'} = (101\pm 23)^\circ$ and $\phi_{B'} =
(145^{+41}_{-27})^\circ$, respectively. The total $\chi^2/({\rm dof}) =
1.64/2=0.82$ compared to the 0.40 of the first choice..

In this case, the fitted values obtained for the phases $\phi_{A'}$ and
$\phi_{B'}$ are compatible with $\pi/2$ and $\pi$, respectively. The
corresponding results obtained for this particular choice in this seven free
parameters fit (the six eigenvalues and $|D'|$) and two degrees of freedom at
the same $M_Z$ energy scale are: $\lambda_{\rm u} =
(1.70^{+0.40}_{-0.36})\,{\rm MeV}$, $\lambda_{\rm c} = -(0.591\pm 0.082)\,{\rm
GeV}$, $\lambda_{\rm t} = (172.5\pm 3.0)\,{\rm GeV}$, $\lambda_{\rm d} =
(2.89\pm 0.69)\,{\rm MeV}$, $\lambda_{\rm s} = -(56\pm 13)\,{\rm MeV}$,
$\lambda_{\rm b} = (2.889\pm 0.090)\,{\rm GeV}$, and $|D'| = (19\pm 13)\,{\rm
MeV}$. The total $\chi^2/({\rm dof}) = 2.72/4=0.68$, very similar to the
corresponding 0.66 of the first choice.

\section{\label{sectioncgs}CGS type mass matrices}

CGS-type hermitian mass matrix was considered
first in Refs.~\cite{chaturvedi08,gupta09} and is
given by

\begin{equation}
M=
\begin{pmatrix}
0 & a & d \\
a^* & 0 & b \\
d^* & b^* & c
\end{pmatrix}
.
\label{cgs}
\end{equation}

\noindent
For $d = 0$ this reduces to the Fritzsch type. This mass matrix
has the virtue that it will give CP-violation in all the three
bases because ${\rm
Im}(M_{12}\,M_{23}\,M^{*}_{13})\ne0$~\cite{chaturvedi08}. To
reduce the number of parameters to four we can take $a$, $b$,
and $c$ to be real and with $d$ as pure imaginary.

In the up (down) quark diagonal basis $M = M_{\rm d}$ ($M_{\rm
u}$) there are seven parameters, 4 in $M$ and 3 masses in the
diagonal mass matrix.

In the physical basis with the CGS-type mass matrix it is enough
to consider the case where $M_{\rm u}$ is Fritzsch-type while
$M_{\rm d}$ is CGS-type~\cite{chaturvedi08,gupta09}

\begin{equation}
M_{\rm u}=
\begin{pmatrix}
0 & a & 0 \\
a & 0 & b \\
0 & b & c
\end{pmatrix}
,
\quad\quad
M_{\rm d}=
\begin{pmatrix}
0 & a' & i|d'| \\
a' & 0 & b' \\
-i|d'| & b' & c'
\end{pmatrix}
.
\label{fritzschcgs}
\end{equation}

\noindent
All the matrix elements are taken to be real and positive, so we
have seven real parameters. This is a small (but important)
variation of Eq.~(\ref{fritzsch}) above.

There exists a unitary matrix $Y$ such that it can rotate $M_u$ and $M_d$ in
Eq.~(\ref{fritzschcgs}) to mass matrices $M'_u = Y^\dag M_uY$ and $M'_d =
Y^\dag M_dY$ which are CGS-type and Fritzsch-type respectively. Explicitly, the
unitary matrix

\begin{equation}
Y = \frac{1}{b}
\begin{pmatrix}
\beta & \delta & 0 \\
\delta & \beta & 0 \\
0 & 0 & b
\end{pmatrix}
,
\end{equation}

\noindent
where $b^2 = \beta^2 + |\delta|^2$ and $\delta =
i\eta_{\delta}|\delta|$ $(\eta_{\delta} = \pm1)$ is pure imaginary. This gives,

\begin{equation}
M'_u =
\begin{pmatrix}
0 & a & \delta^* \\
a & 0 & \beta \\
\delta & \beta & c
\end{pmatrix}
,
\quad\quad
M'_d =
\begin{pmatrix}
0 & a' & \delta' \\
a' & 0 & \beta' \\
-\delta' & \beta' & c'
\end{pmatrix}
.
\end{equation}

\noindent
The conditions which make $Y$ a unitary matrix ensures $M'_u$ and
$M_u$ have the same eigenvalues. Further, in $M'_d$, $b\beta' = d'\delta^* +
\beta b'$ and $b\delta' = \beta d' + b'\delta^*$. Since $d'$ and $\delta$ are
pure imaginary so is $\delta'$ while $\beta'$ is real. Thus, both $M'_u$ and
$M'_d$ are of the CGS-type. By requiring, $\delta' = 0$ we can make $M'_d$ to
be Fritzsch-type. Thus, it is sufficient to consider the case $M_u$
Fritzsch-type and $M_d$ CGS-type (see Eq.~(\ref{fritzschcgs})) in this analysis.

From the characteristic polynomial of $M_{\rm u}$ and $M_{\rm d}$ we can
solve for the matrix elements in terms of the corresponding eigenvalues
and $|d'|$,

\[
a=\left[
-\frac{\lambda_{\rm u}\lambda_{\rm c}\lambda_{\rm t}}
{\lambda_{\rm u}+\lambda_{\rm c}+\lambda_{\rm t}
}
\right]^{1/2}
,
\qquad
c=\lambda_{\rm u}+\lambda_{\rm c}+\lambda_{\rm t}\,,
\]

\begin{equation}
b=\left[-
\frac{
(\lambda_{\rm t}+\lambda_{\rm c})
(\lambda_{\rm t}+\lambda_{\rm u})
(\lambda_{\rm c}+\lambda_{\rm u})}
{\lambda_{\rm u}+\lambda_{\rm c}+\lambda_{\rm t}
}
\right]^{1/2}
,
\label{15bnon}
\end{equation}

\noindent
and

\[
a'=\left[
-\frac{\lambda_{\rm d}\lambda_{\rm s}\lambda_{\rm b}}
{\lambda_{\rm d}+\lambda_{\rm s}+\lambda_{\rm b}
}
\right]^{1/2}
,
\qquad
c'=\lambda_{\rm d}+\lambda_{\rm s}+\lambda_{\rm b}\,,
\]

\begin{equation}
b'=\left[-
\frac{
(\lambda_{\rm b}+\lambda_{\rm s})
(\lambda_{\rm b}+\lambda_{\rm d})
(\lambda_{\rm s}+\lambda_{\rm d})}
{\lambda_{\rm d}+\lambda_{\rm s}+\lambda_{\rm b}
} - |d'|^2
\right]^{1/2}
.
\label{15bpnon}
\end{equation}

From Eq.~(\ref{5}), the theoretical expressions for the observables the
magnitudes of the unitary quark mixing matrix elements in this case are

\begin{eqnarray}
|V_{\alpha\,j}|^2&=&
\big{[}(\lambda_{\beta} - \lambda_{\alpha})
(\lambda_{\gamma} - \lambda_{\alpha})
(\lambda_k - \lambda_j)
(\lambda_l - \lambda_j)\big{]}^{-1}\times
\nonumber
\\
&\ &
\big{[}
(a\,b'+b\,a')^2
+ 2(a^2a'^2+b^2b'^2)+(a^2+b^2)|d'|^2
\nonumber
\\
&\ &+\,
2\,a\,a'(\lambda_{\beta} + \lambda_{\gamma})
(\lambda_k + \lambda_l)
+\lambda_{\beta}\lambda_{\gamma}
\lambda_{k} \lambda_l
\nonumber
\\
&\ &+\,
\lambda_{k} \lambda_l
(
a^2+b^2+\lambda^2_{\alpha}
)+
\lambda_{\beta}\lambda_{\gamma}
(
a'^2+b'^2+|d'|^2+\lambda^2_j
)
\nonumber
\\
&\ &+\,
c\,\lambda_{\alpha}(b'^2+|d'|^2)
+c'\lambda_{j}\,b^2
+\lambda_{\alpha}\lambda_{j}(c\,c'+2b\,b')
\big{]}
\,,
\label{vajnon}
\end{eqnarray}

\noindent
where ($\alpha$,$\beta$,$\gamma$) is any permutation of (u,c,t)
and ($j$,$k$,$l$) any permutation of (d,s,b).

The theoretical expression for the Jarslkog invariant $J(V)$ given by
Eq.~(\ref{12}) translates into,

\begin{equation}
J(V) = -
\frac{
b\,|d'|\,\big{[}(a'b - a\,b')\,(b\,c' - b'c) +
a\,c\,|d'|^2\big{]}
}{
(\lambda_{\rm b} - \lambda_{\rm s}) (\lambda_{\rm b} -
\lambda_{\rm d})(\lambda_{\rm s} - \lambda_{\rm d})
(\lambda_{\rm t} - \lambda_{\rm c}) (\lambda_{\rm t} -
\lambda_{\rm u})(\lambda_{\rm c} - \lambda_{\rm u})
}.
\label{jvnon}
\end{equation}

As in the Fritzsch case it is possible to fix the relative signs between the
eigenvalues and the quark masses using the quark mass hierarchies and the
characteristic equations, Eqs.~(\ref{15bnon}) and (\ref{15bpnon}), again in
this case are valid the identifications of Eq.~(\ref{17p}).

The seven free parameters to be estimated are the six eigenvalues ($\lambda_{\rm
u}$,$\lambda_{\rm c}$,$\lambda_{\rm t}$) and ($\lambda_{\rm d}$,$\lambda_{\rm
s}$,$\lambda_{\rm b}$), and the matrix element $|d'|$. The number of degrees of
freedom is four. At the $M_Z$ energy scale the fitted values for these
parameters are: $\lambda_{\rm u} = (1.30^{+0.49}_{-0.43})\,{\rm MeV}$,
$\lambda_{\rm c} = -(0.643\pm 0.081)\,{\rm GeV}$, $\lambda_{\rm t} = (172.4\pm
3.0)\,{\rm GeV}$, $\lambda_{\rm d} = (2.79^{+0.39}_{-0.38})\,{\rm MeV}$,
$\lambda_{\rm s} = -(35.7\pm 4.6)\,{\rm MeV}$, $\lambda_{\rm b} = (2.899\pm
0.090)\,{\rm GeV}$, and $|d'| = (8.3^{+1.5}_{-1.1})\,{\rm MeV}$. The total
$\chi^2/({\rm dof}) = 1.89/4=0.47$.

This is a seven free parameters fit with $M_{\rm u}$ being Fritzsch type while
$M_{\rm d}$ is CGS type with $a'$ and $b'$ real, and $d'$ pure imaginary.

\section{\label{conclusions}Conclusions and discussion of results}

The results for the fits for all quark masses and matrix elements at the energy
scale of $M_Z$ are summarized in column $M_Z$ of Table~\ref{table2}. As can be
seen, with 9 parameters the best fits are obtained using a Gupta-Rajpoot type
matrix, as expected. However, fixing $\phi_{A'} = -\pi /2$ and $\phi_{B'} = 0$
(as suggested by the free fits) in the Fritzsch case, reducing the number of
parameters to only 6, gives a comparable $\chi^2 /{\rm dof}$ than fixing
$\phi_{A'} = -\pi /2$ and $\phi_{B'} = 0$ (also suggested by the free fits and
reducing the number of parameters to 7) in the Gupta-Rajpoot case. With 7
parameters the best ratio $\chi^2 /{\rm dof}$ is obtained using a CGS type
matrix.

The stability of this type of analysis with respect to evolution of the quark
masses is important, in column 2~GeV of Table~\ref{table2} we summarize the
results for the different cases at 2~GeV scale. As can be seen, the results for
$\chi^2 /{\rm dof}$ are very similar to results at $M_Z$ scale.

A simple way to understand this is to note that if all quark masses are scaled
by a common factor then the algebraic expressions for the dimensionless numbers
$J(V)$ and the moduli $|V_{\alpha j}|$ ($\alpha$=u, c, j=d, s) will be
unaffected. As can be seen from Table~\ref{table1}, the ratio $m_{\rm q}
(2~{\rm GeV})/m_{\rm q}(M_Z) = 1.71-1.74$, for q=u, d, s, c, b, while it is
1.85 for q=t.

Finally, column PDG of Table~\ref{table2} gives the results for the different
cases for quark masses at different scales (the PDG convention) given in
column PDG of Table~\ref{table1}. Again, the results are similar and choice of
CGS mass matrices is favored. Note that masses of the heavier quarks (notable
$m_{\rm t}$) are very different. This would suggest that the role of the small
masses (since they evolve slowly) is possibly more important.

In conclusion, we advocate the use of mass matrices which can be used in all the
three bases.

\begin{acknowledgments}

V.~Gupta and G.~S\'anchez-Col\'on would like to thank CONACyT (M\'exico) for
partial support. The work of S.~Rajpoot was supported by DOE Grant \#: {
DE-FG02-10ER41693}.

\end{acknowledgments}

\clearpage

\begin{table}

\caption{Quark masses at various energy scales. Those in columns~2
and 3 are the evolved masses taken from Ref.~\cite{xing08}.
Column~4 gives masses given in Ref.~\cite{pdg06}: $m_{\rm u}$,
$m_{\rm d}$ and $m_{\rm s}$ are at 2~GeV, $m_{\rm c}$ at $m_{\rm
c}$, $m_{\rm b}$ at $m_{\rm b}$ and $m_{\rm t}$ by direct
observation. Note that u, d, and s masses are in MeV while those
of c, b, and t are in GeV. The experimentally observed
values~\cite{pdg06} of the four best measured magnitudes of the
quark mixing matrix elements and the Jarlskog invariant are
displayed in the second part of the table. \label{table1}}

\begin{tabular}{cccc}

\hline
Quark mass & \multicolumn{3}{c}{Scale} \\
 & 2~GeV & $M_Z=91.1876\,{\rm GeV}$ & PDG \\
\hline
$m_{\rm u}$ & $2.2^{+0.8}_{-0.7}$ & $1.28^{+0.50}_{-0.43}$&
$2.25 \pm 0.75$ \\
\multicolumn{4}{c}{} \\
$m_{\rm d}$ & $5.0\pm 2.0$ & $2.91^{+1.24}_{-1.20}$ &
$5.0 \pm 2.0$ \\
\multicolumn{4}{c}{} \\
$m_{\rm s}$ & $95 \pm 25$ & $55^{+16}_{-15}$ &
$95 \pm 25$ \\
\multicolumn{4}{c}{} \\
$m_{\rm c}$ & $1.07 \pm 0.12$ & $0.624 \pm 0.083$ &
$1.25 \pm 0.09$ \\
\multicolumn{4}{c}{} \\
$m_{\rm b}$ & $5.04^{+0.16}_{-0.15}$ & $2.89 \pm 0.09$ &
$4.20 \pm 0.07$ \\
\multicolumn{4}{c}{} \\
$m_{\rm t}$ & $318.9^{+13.1}_{-12.3}$ & $172.5 \pm 3.0$ &
$174.2 \pm 3.3$ \\

\hline
Observable & \multicolumn{3}{c}{Exp. value} \\
\hline
$|V_{\rm ud}|$ & \multicolumn{3}{c}{$0.97383 \pm 0.00024$} \\
$|V_{\rm us}|$ & \multicolumn{3}{c}{$0.2272 \pm 0.0010$} \\
$|V_{\rm cd}|$ & \multicolumn{3}{c}{$0.2271 \pm 0.0010$} \\
$|V_{\rm cs}|$ & \multicolumn{3}{c}{$0.97296 \pm 0.00024$} \\
$J(V)$ & \multicolumn{3}{c}{$(3.08 \pm 0.18)\times 10^{-5}$} \\
\hline

\end{tabular}

\end{table}

\begin{table}

\caption{ Fits for all quark masses and matrix elements at energy scales of
$M_Z$, 2 GeV, and at the PDG energy scales convention. \label{table2}}

\begin{tabular}{c|c|c|c|c|c}

\hline

Type of & {Physical} & Number of & \multicolumn{3}{c}{$\chi^2/({\rm
dof})$} \\ \cline{4-6}

mass matrix & {Basis} & param. & {$M_Z$} & {2 GeV} &
{PDG} \\

\hline

{Fritzsch} & $\phi_{A'}$ and $\phi_{B'}$ free. & 8 &
4.23/3 = 1.41 & 4.80/3 = 1.60 & 3.32/3 = 1.11 \\

& $\phi_{A'} = -\pi/2$, $\phi_{B'} = 0$. &6 & 4.84/5= 0.97 & 5.49/5= 1.10 &
4.27/5= 0.85 \\

\hline

Gupta-Rajpoot & $\phi_{A'}$ and $\phi_{B'}$ free. & 9 & 0.80/2 = 0.40 &
0.87/2 = 0.44 & 0.78/2 = 0.39 \\

$|D'|=0$, $D\ne 0$. & $\phi_{A'} = -\pi/2$, $\phi_{B'} = 0$. & 7 & 2.62/4= 0.66
& 2.76/4= 0.69 & 3.77/4= 0.94 \\

\hline

Gupta-Rajpoot & $\phi_{A'}$ and $\phi_{B'}$ free. & 9 & 1.64/2 = 0.82 &
1.77/2 = 0.89 & 1.71/2 = 0.86 \\

$|D'|\ne 0$, $D=0$. & $\phi_{A'} = \pi/2$, $\phi_{B'} = \pi$. & 7 & 2.72/4=
0.68 & 2.86/4= 0.72 & 3.83/4= 0.96 \\

\hline

$M_{\rm u}$ Fritzsch-type &  & {7} & {1.89/4 = 0.47 } & 2.47/4 = 0.62 &
0.80/4 = 0.20 \\

and $M_{\rm d}$ CGS-type. & & & & & \\

\hline

\end{tabular}

\end{table}

\end{document}